# *DETECTING DISTRIBUTED DENIAL OF SERVICE ATTACKS USING LOGISTIC REGRESSION AND SVM METHODS*

Mohammad Arafat Ullah
*Computer Science and Engineering*
*East West University*
Dhaka, Bangladesh
arafat.ngc@gmail.com

Arthy Anjum Jamal
*Computer Science and Engineering*
*East West University*
Dhaka, Bangladesh
arthy.anjum@gmail.com

Rashedul Amin Tuhin
*Computer Science and Engineering*
*East West University*
Dhaka, Bangladesh
mcctuhin@ewubd.edu

Shamim Akhter
*Computer Science and Engineering*
*International University of Business Agriculture and Technology (IUBAT)*
shamimakhter@gmail.com

***Abstract*—A distributed denial-of-service (DDoS) attack is an attempt to produce humongous traffic within a network by overwhelming a targeted server or its neighboring infrastructure with a flood of service requests ceaselessly coming from multiple remotely controlled malware-infected computers or network-connected devices. Thus, exploring DDoS attacks by recognizing their functionalities and differentiating them from normal traffic services are the primary concerns of network security issues particularly for online businesses. In modern networks, most DDoS attacks occur in the network and application layer including HTTP flood, UDP flood, SIDDOS, SMURF, SNMP flood, IP NULL, etc. The goal of this paper is to detect DDoS attacks from all service requests and classify them according to DDoS classes. In this regard, a standard dataset is collected from the internet which contains several network-related attributes and their corresponding DDoS attack class name. Two(2) different machine learning approaches, SVM and Logistic Regression, are implemented in the dataset for detecting and classifying DDoS attacks, and a comparative study is accomplished among them in terms of accuracy, precision, and recall rates. Logistic Regression and SVM both achieve 98.65% classification accuracy which is the highest achieved accuracy among other previous experiments with the same dataset.**

## I. INTRODUCTION

An evil-intentioned attempt to turmoil in the normal flood of a destination server by overflowing the destination through a flood of Internet traffic is called a denial of service attack (DoS). When multiple compromised computers or IoT devices are used as a source of attack to accomplish the potency of DDoS attacks then it is called distributed denial of service attack. A DDoS attack is similar to a traffic jam on the highway, preventing traffic from arriving at its desired location, consequently, a DDoS attack prevents the normal flood of a networking system from arriving at its destination server. The major motive of a DDoS attack is to disturb and unavailable the normal service provided by a site for both internal and external users. New types of DDoS attacks are generated day by day by hackers who work both the application layer and network layer. Attackers have found the exposures in the application layer to attack the victim and using DDoS makes it more complex in finding and handling the attack.

It is a complex procedure to detect DDoS manually. A security engineer works continuously to monitor and detect packets of DDoS attacks. These are very challenging tasks and the accuracy level depends on human effectiveness. Intrusion detection System (IDS) is another technique used to detect DDoS. IDS system uses machine learning to detect or classify network traffic as normal or as a type of DDoS based on some attributes including average packet size, inter-arrival time, packet size, packet rate, bit rate, etc. Usually, DDoS attacks have the same average packet size, frequently requested packets, increased number of packets, high bit rates packets, and other attributes that help to consume network bandwidth and make services unavailable to users.

We are proposing two (2) new machine learning approaches to detect and classify the modern type of distributed denial of service attacks like HTTP flood and SIDDOS according to their given 27 attributes. Later, we compare ML models based on their performance on classification accuracy.

The remainder of the paper is organized as follows: Section II presents the related work and provides a brief discussion of machine learning classifiers in the relevant area. In Section III, we present the overview of the used data set. Description of the attributes and type of the attacks are also presented in this section. Section IV describes the proposed methodology of implementation and used machine learning approaches. In section V, we describe the evaluation metrics used to assess the performance of the classifiers with the details of the experimental setup and achieved results. Finally, we conclude the paper and describe the future work in section VI.

## II. RELATED WORK

Several approaches have been proposed so far in the case of DDoS Detection. Those approaches are generally categorized into two types. One belongs to signature based the other belongs to anomaly-based. In the case of a signature-based approach disrupted network flood can be detected by checking previous anomaly records. Hence if a new attack arrives it cannot be detected through using a signature. It can be detected through anomaly-based intrusion detection which includes machine learning, analyzing time intervals, checking deviation of IP count, etc. Sometimes an attack may be provided through sending duplicate requests. It can be detected and also be prevented by analyzing the cookies of the attackers if and only if those cookies were previously stored in the victim's server which is discussed in this paper[1]. It is a signature-based intrusion detection system. It includes analyzing the hash values of cookies and modifying them. This system lacked in detecting those attacks in which the producer's cookies have not been stored in the victim's server already.

Then machine learning has come into play to classify the behavior of network traffic by analyzing time intervals between two consecutive network traffic[2]. However, only the time intervals feature is not enough to detect modern types of DDoS. Next, a new approach was provided in this paper[3] which mitigates botnet activity by using session keys and providing CAPTCHA.

In January 2015 a new paper[4] was published regarding DDoS Detection which includes analyzing delta time between two consecutive HTTP requests to detect attacks in the $7^{th}$ layer of the OSI model or the application layer. The smaller the delta time the bigger will be the possibility of the attack. Then some machine learning algorithms Naïve Bayes, Naïve Bayes Multinomial, Multilayer Perceptron, RBF Network, Random Forest, and Logistic Regression have been implemented to the obtained features. Naïve Bayes Multinomial has achieved the highest accuracy rate of 93.67%. The major drawback of this research was that it used a medium-scale attack scenario with only 1000 clients and a web server which was not sufficient for detecting real-time DDoS attacks. A new dataset was collected in this paper[5] which contains new types of DDoS attacks like HTTP flood and SIDDOS. Then three different classification algorithms were implemented which were Multilayer Perceptron, Naïve Bayes, and Random Forest. Multilayer Perceptron achieved the highest accuracy rate of 98.63%. To implement these classification approaches it has used a machine learning tool called Weka version 3.7.12. It also achieved the highest precision and recall rate. Though these accuracy, precision, and recall rates are enough to define a model as an appropriate model a comparison study was necessary with other classifications and other hybrid approaches to verify its perfectness more clearly.

The research work by Loukas et al[6] achieved some level of success if the r-RNN was trained very well. However, one major drawback to this technique is that the results were tested only on a standalone outdated dataset that did not have current DDoS attack features. Sabrina et al[6] detection procedure was based on the use of the human brain to manually observe and make assumptions, and the automated way of using Artificial intelligence (RNN Ensemble). The drawback of this approach pertains to the observation stage where one assumes that the increase in the number of rejected requests and the increase in system resources indicates that the victim is under attack. This is because such incidents can also occur as a result of normal user usage.

In October 2017 a new approach was proposed for detecting DDoS in which the C.4.5 algorithm was coupled with the signature-based (SNORT) technique in this paper[7]. This was implemented in a sample dataset that contains five parameters which are Protocol, Flag, Service, and TTL (Time to Leave). Those parameters were necessary for detecting layer 3 and layer 4 attacks of the OSI model. The implemented C.4.5 model achieved 98.8% with 0.58 seconds of training time. Then a Naïve Bayes and a K-mean approach was implemented for a comparison study. Among them, the C.4.5 approach provided better performance. The major drawback of this system was that this research did not use real-time attack traffic. Hence it cannot be told what performance it will provide in the case of real data.

DDoS attack detection by using the Unique Source IP deviation method was proposed in this paper[8]. The name of the detection method was Violating Source IP Count(VSC) which includes counting unique source IP addresses and the number of violating source IP addresses and calculating their deviations from their mean. If the deviation value is too high then the system is under attack. It used the Non-Parametric CUSUM algorithm for implementing the above theory.

In the next paper[9] deep neural network was implemented with NSL-KDD data in two different ways. In the first one, the output class is considered a binary class( DDoS class and Normal class). The next experiment was performed output class as multiclass considering all types of cyber-attacks differently. The first one provided 98.8% accuracy and the next one provided 94.8% accuracy. NSL-KDD has not contained modern types of DDoS attacks like HTTP flood and SIDDOS. Also, the system lacks in the case of using hybrid machine learning approaches as well as tuning deep neural network parameters.

In August 2018 Animesh Gupta[10] used KDD Cup 1999 dataset for detecting DDoS. Since this dataset contains many features WRFS algorithm was used for selecting necessary features. Then Naïve Bayes, Decision Tree, Random Forest, and SVM four different classification algorithms were implemented in extracted features. But in this approach, the output class is considered as a binary class only normal class and DDoS class though this dataset contains several types of DDoS classes like back-dos, buffer-overflow u2r, ftp-write r21, guess-password, SMURF dos, etc. There was a major lack of this approach.

Three (3) research directions are found related to the DDoS attacks problem. The first is to trace the new features/attributes like duplicate requests, not matching previously stored hash values of cookies, the time interval between two consecutive network traffics, session keys and CAPTCHA values, delta time between two consecutive HTTP requests, etc. The second is the number of DDoS type detection or classification including binary to multiple classes. The third is to use different machine-learning approaches to improve the classification accuracy. In our proposed schemes we are highlighting all three (3) factors together and thus choosing a multi-attributes and multi-classes data set, which is used in [5]. Multilayer Perceptron, Naïve Bayes and
Random Forest were the machine learning tools used to find the classification accuracy. However, we are proposing two (2) additional classifiers which are not yet implemented on this data set, and thus their accuracy is still missing. Our main objective is to implement different classifiers including Logistic Regression and SVM approaches to improve the accuracy of other previously experimented models on this particular data set.

## III. Dataset Overview

### *A. Data Collection*

The distributed denial of service data has been collected from this paper[5]. The major privilege of this dataset is it contains data about modern types of DDoS like SIDDOS and HTTP flood. Moreover, the dataset is very much clean and clear, it does not contain any duplicate or redundant records.

### *B. Dataset Description*

The dataset contains 27 attributes as well as a class-level attack name. It contains 1048574number of records with five different classes. Each of the classes corresponds to a type of DDOS flood except the normal flood. The names and types of the 27 attributes are shown in TABLE I.

TABLE I. All Features of the Collected Dataset

| Variable No | Description | Type |
|---|---|---|
| 1 | SRC_ADD | Continuous |
| 2 | DES_ADD | Continuous |
| 3 | PKT_ID | Continuous |
| 4 | FROM_NODE | Continuous |
| 5 | TO_NODE | Continuous |
| 6 | PKT_TYPE | Continuous |
| 7 | PKT_SIZE | Continuous |
| 8 | FLAGS | Symbolic |
| 9 | FID | Continuous |
| 10 | SEQ_NUMBER | Continuous |
| 11 | NUMBER_OF_PKT | Continuous |
| 12 | NUMBER_OF_BYTE | Continuous |
| 13 | NODE_NAME_FROM | Symbolic |
| 14 | NODE_NAME_TO | Symbolic |
| 15 | PKT_IN | Continuous |
| 16 | PKT_OUT | Continuous |
| 17 | PKTR | Continuous |
| 18 | PKT_DELAY_NODE | Continuous |
| 19 | PKT_RATE | Continuous |
| 20 | BYTE_RATE | Continuous |
| 21 | PKT_AVG_SIZE | Continuous |
| 22 | UTILIZATION | Continuous |
| 23 | PKT_DELAY | Continuous |
| 24 | PKT_SEND_TIME | Continuous |
| 25 | PKT_RESERVED_TIME | Continuous |
| 26 | FIRST_PKT_SENT | Continuous |
| 27 | LAST_PKT_RESERVED | Continuous |

### *C. Detail Description of Each Class*

Among those 5 classes, 4 classes belong to the DDoS class rest is the normal class. Detail descriptions of four DDoS classes are given below:

*1) UDP FLOOD:* User Datagram Protocol (UDP) Flood is a network layer DDoS attack. Since it is a connectionless protocol, attackers use other computers or IOT devices as botnets through a command to use network workstations for the completion of the attack. Then he/she could send a huge amount of UDP traffic to the victim's server through the compromised computers or IOT devices of all the workstations. Those affected by the UDP flood provide denial of accepting the server's service. Fig. 1 shows a UDP flood coming from five different sources to attack a server.

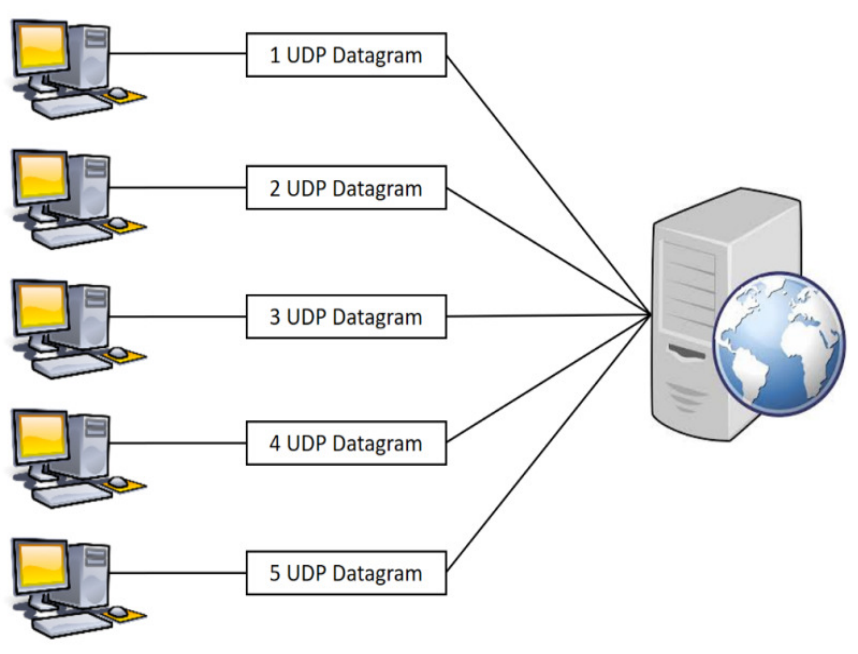

Fig. 1. UDP Flood Structure

*2) SMURF:* It is also a type of denial of service attack that is provided in the network layer. Smurf is provided through cracking IP addresses in a network for a large number of Internet Control Message Protocol (ICMP) traffic to the victim's server. Fig. 2 shows how an attacker can provide ICMP traffic to the victim's server.

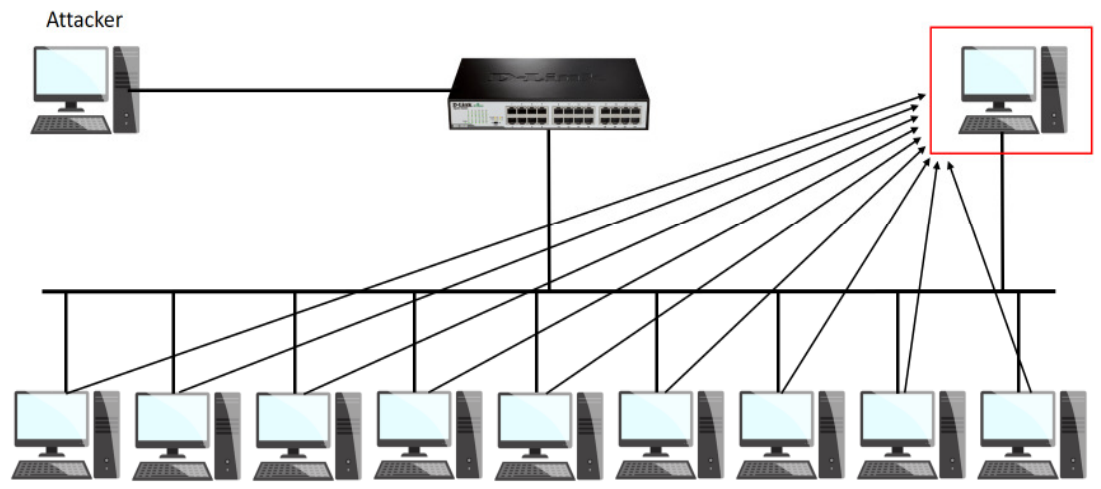

Fig. 2. SMURF Attack Structure

*3)* SIDDOS: SQL Injection DDoS (SIDDOS) is a DDoS attack that occurs in the application layer. This attack is conducted through the insertion of an affected SQL statement as an array and passing this through the database of a website as an equation. Later this illegitimate allowing of this malicious resource provides denial to access the server. Fig. 3 shows how SQL Injection DDoS attacks can bypass the IDS system and attack the webserver followed by the database server.

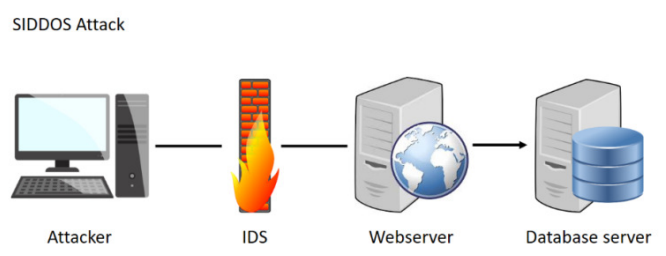

Fig. 3. SIDDOS Attack Structure

*4) HTTP FLOOD:* HTTP FLOOD also occurs in the application layer. Unlike SIDDOS attackers send valid messages instead of illegitimate messages but at a very slow rate to a web application container web server. This provides denial of accessing the server's service through slowing down the server. Flexibleness of accessing the http protocol as well as since it does not provide any illegitimate messages detecting this type of attack is very tough.

### D. Occurrence of Each Class

In this dataset from 1048574 records 939648 belong to normal class which is a rate of 90% of the whole observations. The rest 10% belong to the DDoS class. From that 10%, 9% belong to UDP flood which contains 97521 records from all the observations. The SMURF, SIDDOS, and HTTPS are 6211, 3198, and 1997 of the whole observations respectively which are the rate of 1% and close to 0% respectively. Fig. 4 shows that 90% of the whole dataset contains normal class and the remaining 10% contains four types of DDoS class in which 9% belong to UDP flood, and only 1% belong to the rest of the classes.

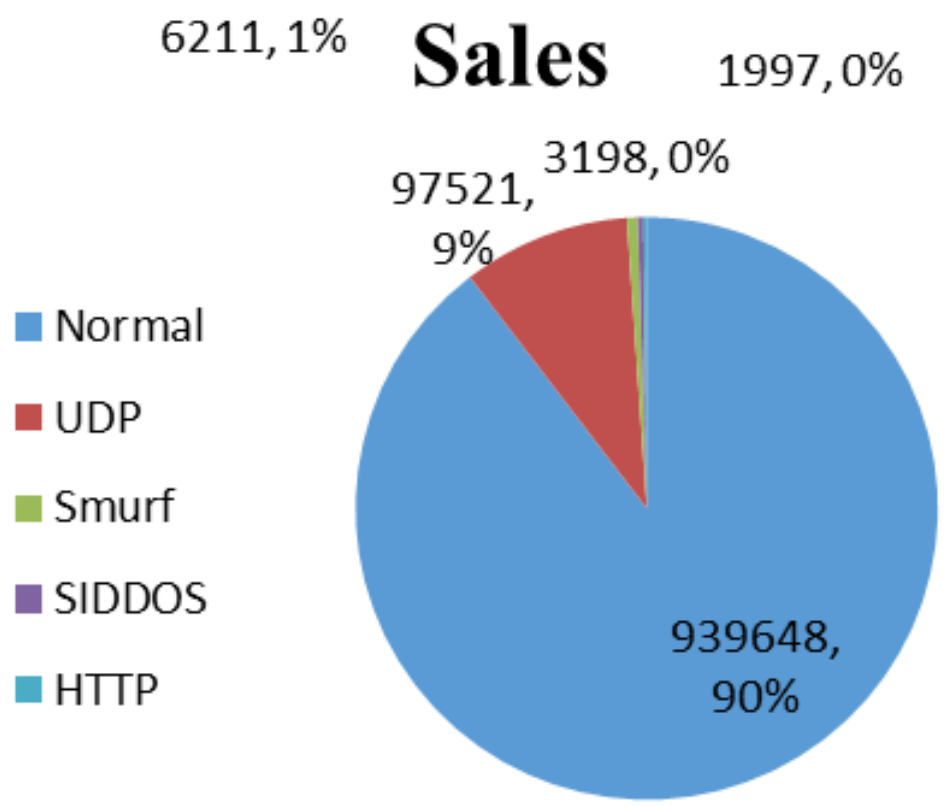

Fig. 4. Pie Chart of the occurrence of each class

## IV. Methodology

At first, the dataset was read in the Linux Mint platform through a machine learning tool Weka version 3.8.3 to visualize the dataset. All dataset information has been obtained through Weka.

### A. Data Preprocessing and Feature Extraction

Machine Learning requires preprocessing of data before fitting any model since machines are only capable of understanding numerical data. The preprocessing steps are given below:

- Our collected dataset was saved in comma-separated value (CSV) format. At first, the dataset was read then it was converted into a data frame through Panda Library for clear visualization and analysis.
- Then a checking operation has been performed for whether it contains duplicated records or not. Any redundant or duplicate records have not been found.
- In the next step, a feature matrix X was conducted which took into account all the features except the class level, and another vector Y was also conducted with this class level.
- Then the indexes of the feature matrix X and the values of vector Y which contain categorical values have been converted into numerical values through label encoding which is a Python library for performing this operation.
- Then feature matrix X has been rescaled through min-max normalization for reducing the feature range for clear observation. The formula of min-max normalization is as follows:

  $$x' = \frac{x - min}{max - min} * (newmax - newmin) + newmin$$

  Here $x'$ =converted value of a given $x$ into a new feature range. $x =$ Initial value of x in previous feature range. Min-Max normalization has been implemented in the feature range of 0 to 1.
- After completion of the feature extraction task, the feature matrix along with the feature vector has been split as 80% train set and 20% test set through model selection.
- Then two different machine learning approaches were implemented which were Logistic Regression, and SVM.

### B. Logistic Regression

Logistic Regression is a statistical analyzing technique in which dependency between two variables has been calculated. The Logistic Regression was initially developed by Joseph Berkson who considered this model as a statistical model [6]. In this research, Logistic Regression has been implemented through the SKLEARN library in Python with an LBSFGS solver since the dataset contains a multiclass level.

### C. SVM

Support vector machine (Also Support Vector Networks [7]) is a popular machine learning algorithm for classification and regression. SVM is applied to DDoS attack detection with good accuracy. Support vector Machine has been implemented through SCIKIT learn with SVC function which contains gamma, decision function shape parameters. In this case, gamma = scale and the ovo decision function shape have been used.

## V. RESULT AND DISCUSSION

### A. *Logistic Regression Model*

*1) Accuracy:* The Logistic Regression model achieved a classification accuracy of 98.65%. Accuracy is the measurement of the rate of the correctly classified instances. Since we have more than two classes in our dataset it cannot tell whether or not all classes are being predicted equally, one or two classes have been avoided.

*2) Confusion Matrix:* A confusion matrix is a table for summarizing the performance of a classification algorithm. The constructed confusion matrix of the test dataset through the Logistic Regression model is presented in Table II.

TABLE II. Confusion Matrix of the Logistic Regression Model

| | | Predicted | | | | |
|---|---|---|---|---|---|---|
| | | Normal | UDP-Flood | Smurf | SID DOS | HTTP-Flood |
| Actual | Normal | 359 | 0 | 21 | 0 | 0 |
| | UDP-Flood | 5 | 187820 | 23 | 0 | 0 |
| | Smurf | 0 | 40 | 567 | 0 | 0 |
| | SID DOS | 6 | 800 | 33 | 404 | 0 |
| | HTTP-Flood | 0 | 1899 | 0 | 0 | 17738 |

The above confusion matrix table shows that this model has taken all classes into account it has not avoided one or two classes. Hence this is a good model.

*3) Precision:* It is the fraction of relevant instances among the retrieved instances.

*4) Recall:* It is the fraction of relevant instances that have been retrieved over the total amount of relevant instances. The classification report which contains the information about precision and recall rate is presented in Table III.

*5) Classification Report:* Almost all of the classes provide high precision and high recall hence the model is very good and effective. It also implies that the input features and output class level have a continuous relationship.

TABLE III. Classification Report of the Logistic Regression Model

| | Precision | recall | F1-score | support |
|---|---|---|---|---|
| Normal | 0.97 | 0.94 | 0.96 | 380 |
| UDP-Flood | 0.99 | 1.00 | 0.99 | 187848 |
| Smurf | 0.88 | 0.93 | 0.91 | 607 |
| SIDDOS | 1.00 | 0.33 | 0.49 | 1243 |
| HTTP-Flood | 1.00 | 0.90 | 0.95 | 19637 |

The classification table (TABLE III) shows that more than 90% F1-score which is a weighted average of the precision and recall rate for all classes except for the SIDDOS class. Hence it can be said that it can classify almost all of the classes except the SIDDOS class. It has also provided support value of each class which is the value of test instances of each class.

### B. *SVM Model*

*1) Accuracy: The* SVM model has provided 98.65% accuracy which is the highest accuracy along with the Logistic Regression model of all the constructed models. The confusion matrix of this model is presented in Table IV. The classification report of this model is presented in Table V.

*2) Confusion Matrix:* The confusion matrix table (TABLE IV) shows that it has equally predicted all of the classes and it has not avoided one or two classes.

TABLE IV. Confusion Matrix of the SVM model

| | | Predicted | | | | |
|---|---|---|---|---|---|---|
| | | Normal | UDP-Flood | Smurf | SIDDOS | HTTP-Flood |
| Actual | Normal | 359 | 0 | 21 | 0 | 0 |
| | UDP-Flood | 5 | 187820 | 23 | 0 | 0 |
| | Smurf | 0 | 40 | 567 | 0 | 0 |
| | SIDDOS | 6 | 800 | 33 | 404 | 0 |
| | HTTP-Flood | 0 | 1899 | 0 | 0 | 17738 |

*3) Classification Report:* All of the classes except the SIDDOS class have achieved high precision and high recall. Only the SIDDOS class provides high precision but low recall. The table (TABLE V) shows SVM also provides almost the same precision and recall rate as the previously used logistic regression model. This model also provides a low F-1 score value for the SIDDOS class which was only 49% and more than 90% for all other classes.

TABLE V. Classification Report of the SVM model

| | precision | recall | F1-score | support |
|---|---|---|---|---|
| Normal | 0.97 | 0.94 | 0.96 | 380 |
| UDP-Flood | 0.99 | 1.00 | 0.99 | 187848 |
| Smurf | 0.88 | 0.93 | 0.91 | 607 |
| SIDDOS | 1.00 | 0.33 | 0.49 | 1243 |
| HTTP-Flood | 1.00 | 0.90 | 0.95 | 19637 |

### *C. Performance Evaluation*

The Logistic Regression model and SVM model have provided good performance in the case of accuracy, precision, and recall. Both models provide similar accuracy of 98.65% and almost similar precision and recall for all of the classes which is the highest achieved performance among other previous models experimented with the same dataset. The previously used Naïve Bayes, Random Forest, and MLP models have achieved 96.91%, 98.02%, and 98.63% respectively. Only the MLP model achieved nearly similar accuracy as our models among the three (3) previous models. The precision and recall rate of previously used models are not experimented with in the previous paper [5]. The comparison graph of accuracy, precision, and recall between SVM and logistic regression is given in Fig.5.

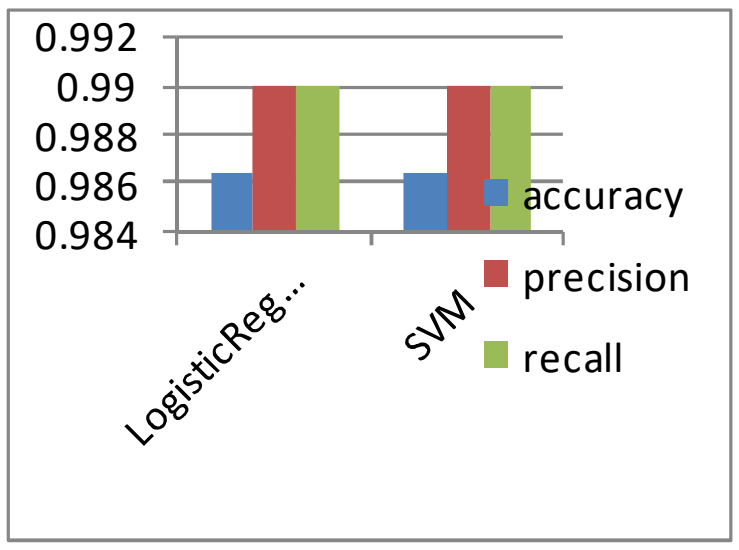

Fig. 5. Accuracy, precision, and recall graph of proposed models

Fig.5 shows that both models provide similar accuracy, precision, and recall rates. Here, weighted precision and weighted recall values of all classes are shown as precision and recall values.

## VI. CONCLUSION

This analysis was about to explore various machine learning approaches for detecting distributed denial of service attacks. A dataset has been collected which contains DDoS types of attack. The collected dataset was a multiclass supervised classification dataset. A comparative study has been performed among all the approaches. In our proposed model the logistic regression model along with the SVM model has achieved 98.65% classification accuracy. Soon, we will add more datasets and combine them with our dataset to classify more new types of DDoS attacks.